\documentclass[twocolumn,superscriptaddress]{revtex4}

\usepackage{amsmath,amssymb,amsfonts,color,graphicx,dsfont}

\begin{document}

\title{Coexistence  of exponentially many chaotic spin-glass attractors}

\author{Y. Peleg}
\affiliation{Department of Physics, Bar-Ilan University, 52900 Ramat-Gan, Israel}
\author {M. Zigzag}
\affiliation{Department of Physics, Bar-Ilan University, 52900 Ramat-Gan, Israel}
\author {W. Kinzel}
\affiliation
{Institute for Theoretical Physics, University of W\"{u}erzburg, Am Hubland, 97074 W\"{u}erzburg, Germany}
\author {I. Kanter}
\affiliation{Department of Physics, Bar-Ilan University, 52900 Ramat-Gan, Israel}
\email{kanter@biu.ac.il}

\begin{abstract}
A chaotic network of size $N$ with delayed interactions which resembles a pseudo-inverse associative memory neural network is investigated.  For a load $\alpha=P/N<1$, where $P$ stands for the  number of stored patterns, the chaotic network functions as an associative memory of $2P$  attractors with macroscopic  basin of attractions which decrease with $\alpha$. At finite $\alpha$, a chaotic spin glass phase exists, where the number of distinct chaotic attractors scales exponentially with  $N$. Each attractor is characterized by a coexistence of chaotic behavior and freezing of each one of the $N$ chaotic units or freezing with respect to the $P$ patterns. Results are supported by large scale simulations of  networks composed of  Bernoulli map units and Mackey-Glass time delay differential equations.
 \end{abstract}	

\maketitle

Over two decades ago a link between statistical physics and neural network theory was established, where associative memory models were mapped onto magnetic Hamiltonian systems with random and frustrated interactions \cite{1,2}.
The capacity, i.e. the number of stored patterns, was calculated using techniques borrowed from random frustrated magnetic systems and in particular advanced method to solve infinite-range spin-glass models \cite{3}.
In parallel to the progress in interdisciplinary research on random magnetic systems, the theory of nonlinear dynamics developed and recently there have been important steps forward in the understanding of conditions to achieve chaotic networks synchronization \cite{4,5}.

It is reasonable to assume that the two phenomena, spin-glasses and chaos, represent two conflicting trends.
The spin-glass (SG) phenomenon is characterized by the existence of exponential number of meta-stable states, in which the local degrees of freedom (spins) are frozen \cite{3,6,6a}.
On the other hand, chaotic dynamics is characterized such that two nearby trajectories diverge from each other as a consequence of the positive Lyapunov exponent \cite{7}.
Nevertheless, a network of coupled chaotic units can synchronize \cite{4,5} to an identical chaotic trajectory.
In this paper we show that chaotic networks can have a SG phase, characterized by an exponential number of chaotic attractors with the system size, $N$.
Each one of the exponentially many attractors is characterized by a coexistence of chaotic behavior and substantial freezing of each one of the N chaotic units.
We start in section \ref{sec: Model} with the definition of the model for a network of chaotic maps.
The associative memory features of the model are analyzed at section \ref{sec: Mattis} and the spin glass phase is examined at section \ref{sec: Spin Glass States}.
The generalization of the chaotic map networks to time delay differential equations is discussed in section \ref{sec: Continuous Time Systems} and concluding remarks are given at section \ref{sec: Conclusions}.

\section{Model}\label{sec: Model}
One of the first examined neural network models within the statistical mechanics framework  is an associative recall of memory without error known also as pseudo-inverse model \cite{8,9}. The model consists of a fully connected network governed by the Hamiltonian
\begin{equation}\label{Hamiltonian}
H\ =\ -\frac{1}{2}\sum_{i \ne j}^N W_{ij} S_i S_j
\end{equation}
where $S_i=\pm 1$ is an Ising spin and the sum is over all pairs of spins.
The coupling matrix $W_{ij}$ is defined as
\begin{equation}\label{pseudo-inverse}
W_{ij}= \frac{1}{N}\sum_{\mu,\nu=1}^{P}\xi_{i}^{\mu}\left(C^{-1}\right)_{\mu \nu}\xi_{j}^{\nu} \quad ,\quad C_{\mu\nu} =\frac{1}{N}\sum_{i=1}^{N}\xi_{i}^{\mu}\xi_{i}^{\nu}
\end{equation}
where $N$ is the size of the network.
The stored patterns $\xi^{\mu}_i\ (\mu=1\ldots P, \ i=1\cdots N)$ are $P$ vectors of length $N$ and
$\xi^{\mu}_i= \pm 1$ with equal probability.
All the patterns are minima of the Hamiltonian and the pseudo-inverse network operates as an associative memory up to $\alpha=P/N=1$ \cite{9}, which is the upper bound for the capacity of such networks with symmetric interactions.
In the case where the diagonal interactions, $W_{ii}$, are introduced the capacity was found to be $\alpha=0.5$ \cite{8}.
The matrix $W_{ij}$ has $P$ eigenvalues equal to $1$ where the patterns are the corresponding eigenvectors,
\begin{equation}\label{eq: eigenvalues}
\sum_{j=1}^N  W_{ij}\xi_j^{\mu} =\xi_i^{\mu},
\end{equation}
and all other $N-P$ eigenvalues are identically zero.

The counterpart chaotic map version of the pseudo-inverse model is given for unit $i$, for instance, by
\begin{align}\label{dynamics}
X_{i}^{t} &= (1-\epsilon)\mathcal{F}(X_{i}^{t-1})+\epsilon\sum_{j=1}^{N}W_{ij}\mathcal{G}(X_j^{t-\tau})
\end{align}
where $W_{ij}$ is given by \eqref{pseudo-inverse} including the diagonal elements,
$\epsilon$ is the strength of the coupling, $1-\epsilon$ is the weight of the internal dynamics
and $\mathcal{F}(\cdot)$ and $\mathcal{G}(\cdot)$ stand for (non-linear) response functions. For simplicity of the discussion below we mainly present results for the simplest chaotic map, the generalized Bernoulli map \cite{10}
\begin{equation}\label{bernoulli map}
\mathcal{F}(x)=\mathcal{G}(x)=sign(x) \big( (a|x|)\mod 1 \big)
\end{equation}
where $\mathcal{F}(x)\in [-1,1]$ and the map is known to be chaotic for $a>1$. For all simulations exemplified below the selected parameters are $a=1.1,~\tau=40$ and $\alpha=0.5$, unless otherwise mentioned.

The state vector $\textbf{X}^{t}$ of the $N$ units at time $t$ can be written as
\begin{equation}\label{state vector}
\textbf{X}^{t}=\sum_{\mu=1}^{P} b_{\mu}^t \xi_{\mu}  + \delta_{\bot}^{t}
\end{equation}
where $\delta_\bot^t$ represents the orthogonal part of $\textbf{X}^{t}$ to the subspace spanned by the $P (\le N)$ patterns.
Since $W_{ij}$ has zero eigenvalues in directions perpendicular to the subspace spanned by the  patterns,
$\delta_\bot^t$ decays exponentially to zero for the discussed region $a(1-\epsilon)<1$.

\section{Mattis States and Associative memory}\label{sec: Mattis}

\begin{figure}[t]
  \includegraphics[width=0.5\textwidth]{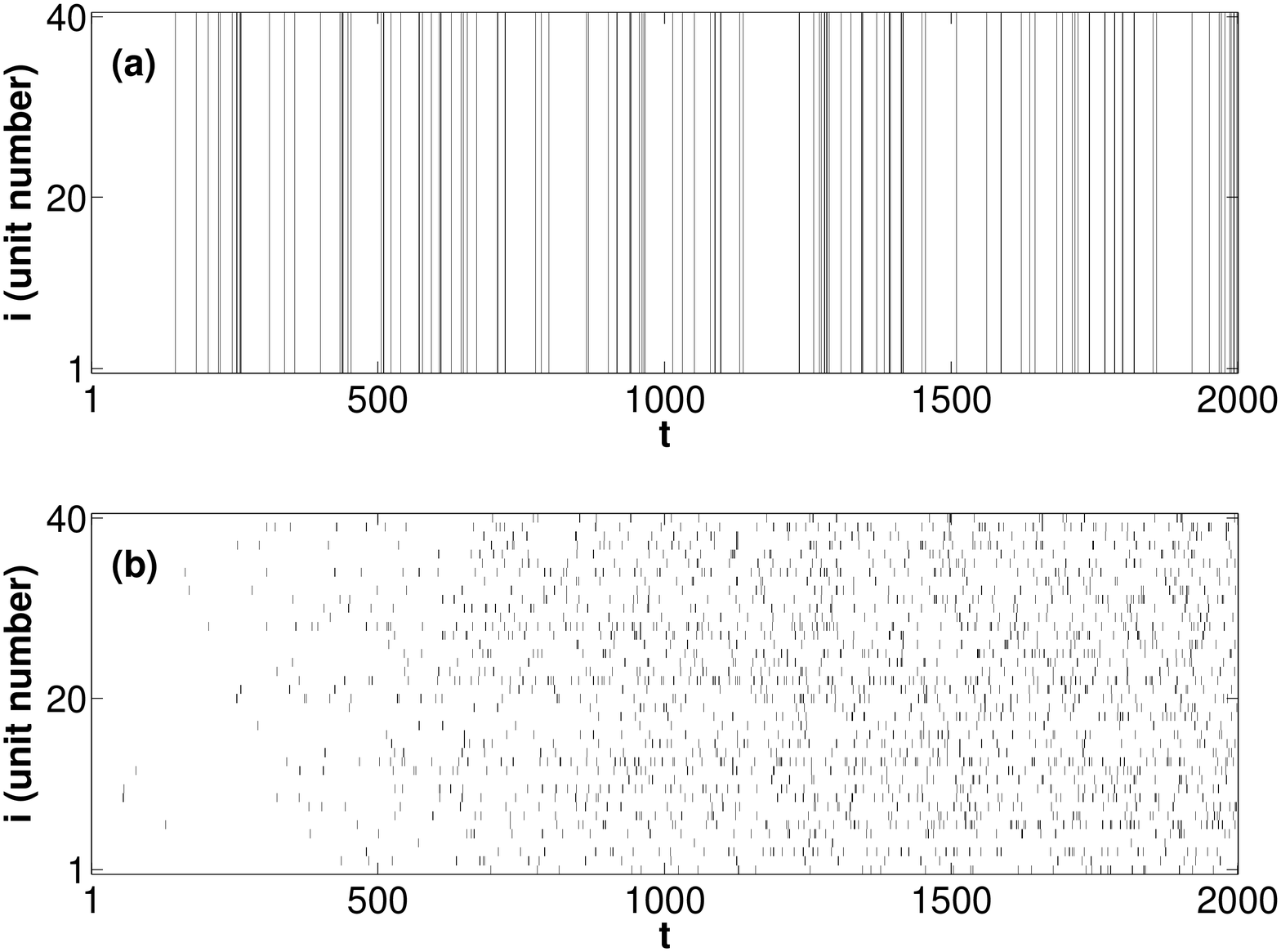}
  \caption{Raster diagrams of the jumps during time evolution of a network ($N=40,P=20$) of chaotic Bernoulli maps, eqs. (\ref{dynamics},\ref{bernoulli map}).
  (a) An initial condition has a unity overlap with one pattern only, eq. \eqref{i.c. one pattern}.
  Jumps occur simultaneously for all units and the overlap with the pattern remains 1.
  (b) Random initial condition. Units jump asynchronously and at a different rate and the maximal overlap with the patterns is less than $3/\sqrt{N}$.
  Note that in case (b) there are more jumps as time evolves since the units get frozen, mostly, in the interval $X_i\in[0.5,1]$ (section \ref{sec: Spin-Glass Phase}).
}\label{fig: raster}
\end{figure}

An  initial condition equals to one of the patterns is a fix point of the Hamiltonian system eqs. (1-2). Nevertheless, for the chaotic dynamics, eqs. (\ref{dynamics},\ref{bernoulli map}), an initial condition which has a unity overlap with pattern $\mu$
\begin{equation}\label{i.c. one pattern}
\textbf{X}^t =b_\mu^t\xi^\mu \quad \left(1 \leqslant t \leqslant \tau ,\ b_\mu^t>0 \right),
\end{equation}
does not lead to a fixed point.
Using the fact that
$\mathcal{F}(b_{\mu}^t\xi_i^{\mu})= \xi_i^{\mu}\mathcal{F}(b_{\mu}^t)$ and
$\sum_{j=1}^N  W_{ij}\xi_j^{\mu} =\xi_i^{\mu}$,
equation \eqref{dynamics} is reduced to
\begin{equation}\label{one pattern}
b_\mu^{t} = (1-\epsilon) \mathcal{F}(b_\mu^{t-1}) +\epsilon \mathcal{F}(b_\mu^{t-\tau})
\end{equation}
which is known to be chaotic for $a>1$ and describes merely the behavior of the coefficient $b_\mu^t$.
For binary patterns, $\xi_i^{\mu} \pm 1$, jumps of the Bernoulli map (i.e. $|X_i^t|>1/a$) occur for all units simultaneously as exemplified by Fig. \ref{fig: raster}a.
The network has has solely a unity projection on pattern $\mu$, and the $N$ dimensional dynamics remains in the direction of the pattern.

For a random  initial condition of the system composed of projections to all patterns,
$\textbf{X}^t =\sum_{\mu=1}^{P} b_\mu^t\xi^\mu \quad \left(1 \leqslant t \leqslant \tau\right)$,
jumps do not occur simultaneously for all units, Fig. \ref{fig: raster}b.
The dynamics of the coefficients $\{ b_{\mu}^t \}$ are coupled and cannot be described independently as in \eqref{one pattern}.
A nontrivial effective coupling among the patterns is generated and time dependent  overlaps with the P patterns is observed.

\begin{figure}[t]
\includegraphics[width=0.5\textwidth]{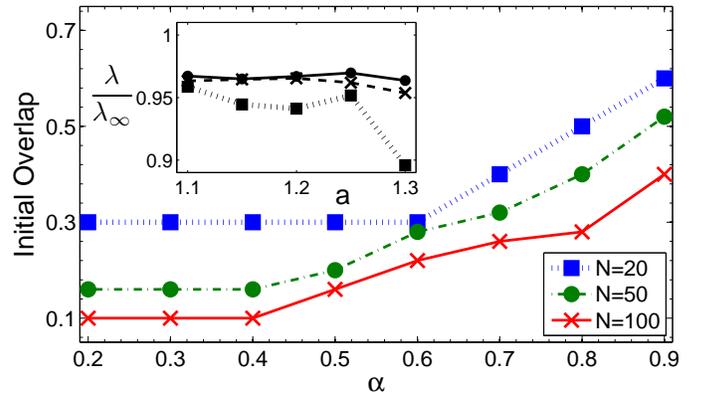}
\caption{(color online). Estimation of the basin of attraction. The average minimal initial overlap with one of the patterns which converges with probability $0.9$ to an overlap $>0.9$ with the same pattern, $N =$ 20($\square$), 50($\circ$), 100($\times$). Simulation parameters are: $\tau = 40, a = 1.1, \epsilon =0.9$, each measured point was averaged over at least  $10^3$ initial overlaps,  the system evolved for $10^3\tau$ steps before  overlaps were averaged over a window of  $\tau$ steps. Inset: Measured Lyapunov exponent, $\lambda$, for both Mattis (an overlap close to unity with one of the pattern) and SG chaotic attractors, normalized to  $\lambda_{\infty}= \tau^{-1} \ln \left| a\epsilon/ (1-a\left(1-\epsilon\right)) \right|$  for $\tau=400$, $N=50$ and each point  averaged over $50$ attractors.}
\label{fig:basin}
\end{figure}

When the initial condition of the system has a projection with one of the patterns which is much larger than all other projections, this  overlap remains the maximal one and furthermore increases towards an overlap close to unity.
Hence, the chaotic pseudo-inverse model \eqref{dynamics} functions as an associative memory with a macroscopic basin of attraction (Fig. \ref{fig:basin}).
Quantitatively, the basin of attraction is defined as the minimal initial overlap with one of the patterns,
and random with the other $N-1$ directions, where the dynamics converges asymptotically to an average overlap greater than $0.9$ with this pattern.
Figure \ref{fig:basin} indicates that the size of the basin of attraction decreases with $\alpha$ and vanishes as $\alpha \rightarrow 1$.
The size of the basin of attraction also increases with the size of the network, $N$, as exemplified in Fig. \ref{fig:basin}.
Similar trends of the basin of attractions have been observed for the Hamiltonian version of the model  \eqref{Hamiltonian}.

Since the dynamics of two  different initial conditions inside the basin of attraction converge to an overlap  close to unity with with the same pattern, it is important to verify whether the dynamics is still chaotic.
The inset of Fig. \ref{fig:basin} presents the Lyapunov exponent, $\lambda$, measured from the divergence of two close initial conditions \textit{within} an attractor for a given set of $(a, \epsilon)$ with $N=400$ and $\tau=50$.
Results indicate a positive Lyapunov exponent
\begin{equation}\label{LE}
\lambda\approx \lambda_\infty=\frac{1}{\tau}\ln \left| \frac{a\epsilon}{1-a\left(1-\epsilon\right)} \right| > 0
\end{equation}
for a large range of  $a(1-\epsilon)<1$,
as was confirmed analytically at for a single chaotic unit with infinite delayed feedback,
$\tau \rightarrow \infty$ \cite{11}.

We exemplify a chaotic network with the coexistence of at least $2P$ chaotic attractors related to the unit eigenvectors,  i.e. the $P$ patterns constructing $W_{ij}$.
Note that the patterns can take any real values instead of the binary patterns in the Ising Hamiltonian, \eqref{Hamiltonian}.
In the next section we examine the existence of other attractors besides the patterns and try to estimate their number.

\section{Spin Glass States}\label{sec: Spin Glass States}

\begin{figure}[t]
\includegraphics[width=0.5\textwidth]{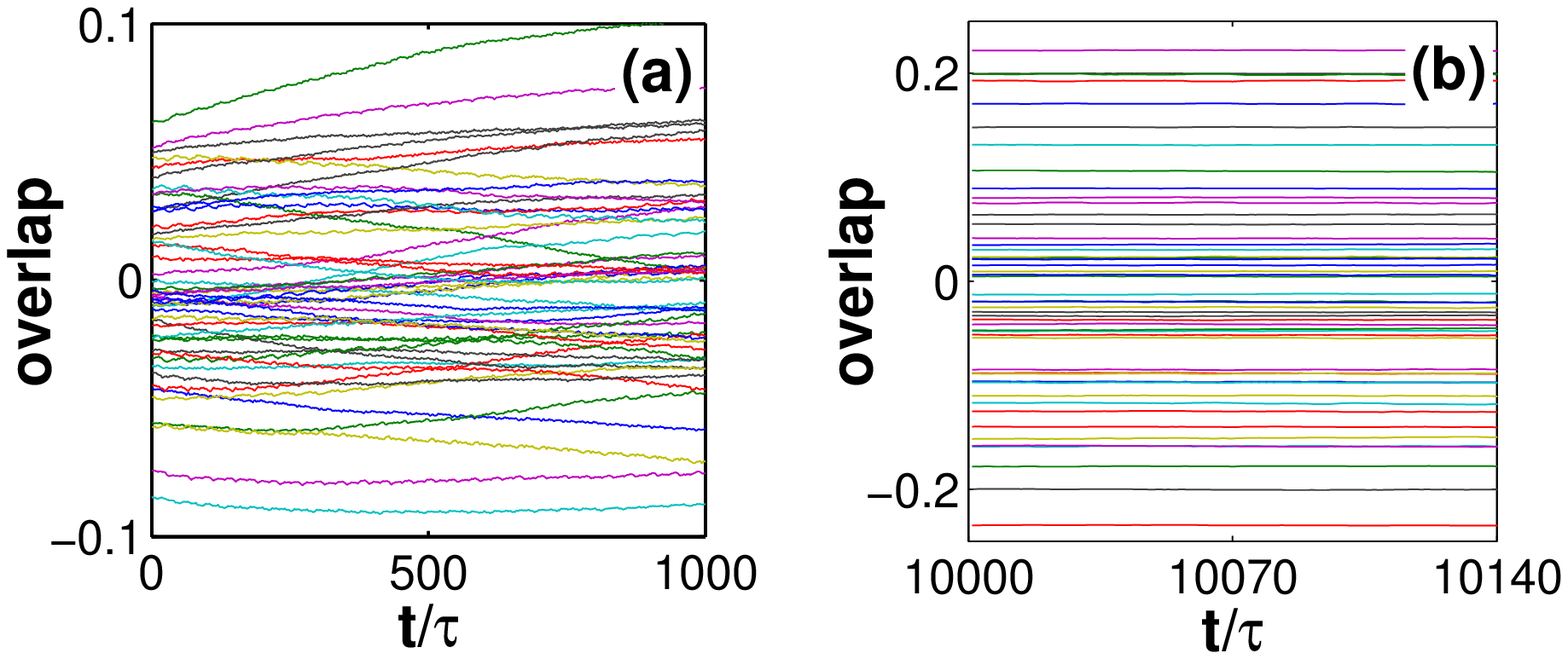}
\includegraphics[width=0.5\textwidth]{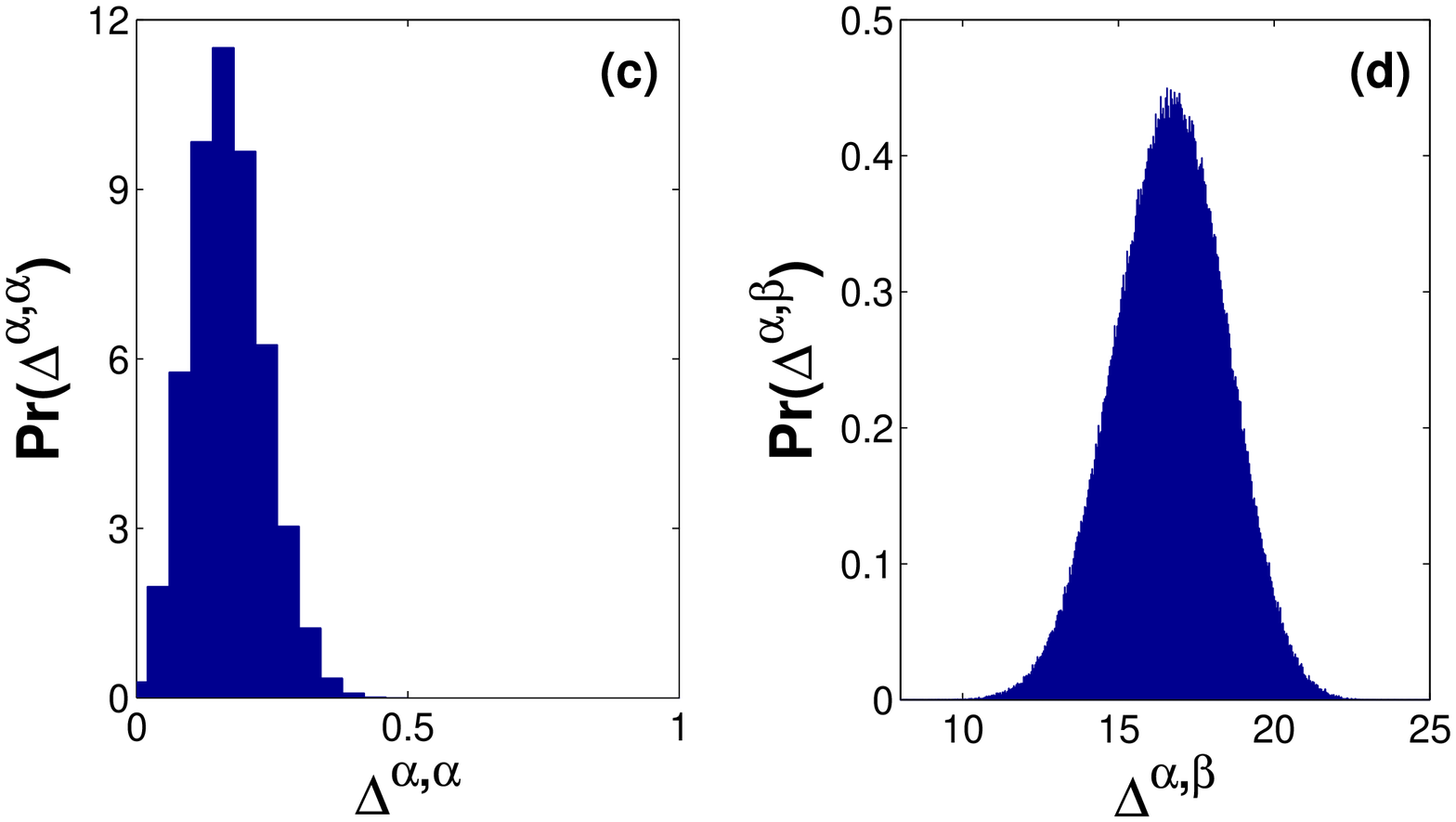}
\caption{(color online). Spin-glass chaotic attractors are exemplified in simulations with $\epsilon=0.9$ and $N = 100$.
(a) The dynamical evolution of the overlaps with the $P$ patterns as a function of time, where each point is averaged over a window of $\tau$ steps. Crossings among the $P$ overlaps are observed during the initial transient time. (b) The overlaps of panel a after the network settle into one of the SG chaotic attractors and the overlaps are constant.
(c) Probability density  of the Manhattan distance along a trajectory while starting from a random initial condition, $\Delta^{\alpha,\alpha}$.The distribution is constructed from the calculation of the averaged $\Delta^{\alpha,\alpha}$  over segments of $10^2\tau$ steps and time offset of  $10^3\tau$ between consecutive segments.
(d) Probability density of the Manhattan distance $\Delta^{\alpha,\beta}$ between the trajectories obtained from two random initial conditions and with  an  offset of $10^4\tau$.  The distribution was constructed similar to panel c.}
\label{fig:hist}
\end{figure}

Starting from random initial conditions, which are typically out of the basin of attraction and consist of overlaps $\sim 1/\sqrt{N}$ with all patterns, result in a SG chaotic attractor.
The dynamical evolution of the $P$ overlaps from a random initial condition up to $10^3\tau$ steps is depicted in Fig. \ref{fig:hist}a, where each point is averaged over a window of $\tau$ steps.
The average is essential since the overlaps dramatically change from one step to another as a consequence of the   discrete time nature of the chaotic map.
The $P$ overlaps varies substantially with time. In addition, crossings among the overlap amplitudes are observed during a transient time.

In contrast, after $10^4\tau$ steps, the average overlaps settled into a well defined order as depicted in Fig. \ref{fig:hist}b.
A similar fixed order in the overlap amplitudes was observed up to $10^6\tau$ steps.
The overlaps are still not macroscopic and are typically bounded (in their absolute value) by $3/\sqrt{N}$ as was confirmed in simulations of network up to $N=1000$ units.
Starting from two close initial conditions within a SG chaotic attractor, a similar Lyapunov exponent $\lambda$  was measured as for Mattis chaotic attractors, eq. \eqref{LE}, independent of $(a,\epsilon)$.
The dynamics are still chaotic, but the trajectory stays (on average) in the neighborhood of a unique direction characterized by a given set of projections with the patterns, Fig. \ref{fig:hist}b.

\subsection{Number of Attractors}\label{sec: Number of Attractors}
Quantitative estimation of the freezing of the $P$ overlaps requires a sophisticated method.
The overlap amplitudes are ranked in a decreasing order.
The pattern with the largest overlap is labeled as $P$, the pattern with the second largest overlap as $P-1$ $\ldots$ and the pattern with the smallest overlap as $1$.
This ranking results in a vector $\textbf{R}$ which is a permutation of $\{1 \ldots P\}$.
Next we introduce the Manhattan distance \cite{10a}
\begin{equation}\label{Manhattan distance}
\Delta^{\alpha,\beta}= \frac{1}{P}\sum_{\mu=1}^{P} |R_\alpha^{\mu}-R_\beta^{\mu}|
\end{equation}
between two rankings, $\alpha$ and $\beta$. Note that $\Delta^{\alpha,\beta} \in [0,P/2]$.
For a random order of $\textbf{R}_\beta$ with respect to $\textbf{R}_\alpha$ the approximate mean value is
\begin{equation}\label{eq: mean delta}
\langle\Delta^{\alpha,\beta}\rangle \approx \frac{1}{3}\left(P-\frac{1}{P}\right),
\end{equation}
where correlations among $R_\alpha^{\mu}-R_\beta^{\mu}$ for different $\mu$'s are neglected.

Results of distribution of the Manhattan distance along different segments of the same trajectory $\alpha$, $\Delta^{\alpha,\alpha}$, when starting from a random initial condition and with an offset time of $10^4\tau$ is depicted in Fig. \ref{fig:hist}c.
The distribution is constructed from the Manhattan distance between each pair of $100$ measurements of $\textbf{R}_\alpha$ with a time offset between two consecutive measurements along the trajectory being $10^3\tau$.
Results indicate an almost complete freezing, where on the average less than a single change in the Manhattan order occurs after an offset of $10^3\tau$ steps.
In order to exclude an accumulated drift in the Manhattan order on much larger time scales, we repeated the distribution of Fig. \ref{fig:hist}c with an offset of $10^4\tau$ and $10^5\tau$ steps among successive measurements and obtained similar distributions.
Figure \ref{fig:hist}d depicts the distribution of $\Delta^{\alpha,\beta}$ between the trajectories of two random initial conditions and with an offset of $10^4\tau$. The most probable distance, $\Delta^{\alpha,\beta}$, is indeed close to the approximated one, $\approx P/3=50/3 \approx 16$, where the lower tail of the distribution ends approximately at $10$.

A comparison between Fig. \ref{fig:hist}c-d for various $N$ indicates that there is a substantial gap between the tails of the distributions  of $\Delta^{\alpha,\alpha}$ and $\Delta^{\alpha,\beta}$ as long as $N\gtrsim30$.
This gap is the fundamental observation which enables us to estimate the number of different SG chaotic attractors.
Assume all SG attractors have similar basin of attractions, one can estimate the number of attractors for a given $N,a,~\epsilon$ and $\alpha$ using the following procedure.
We start sequentially from $N_0$ random initial conditions for the network, eq. \eqref{dynamics}, and after a time offset of at least $10^4\tau$ the stationary ranking $\textbf{R}$ is identified.
In the event the Manhattan distances of this new identified vector with all previously stored ranking vectors are greater than $1$,  a new SG attractor is identified. Next, its ranking vector $\textbf{R}$ is added to the list of stored SG attractors.
In order to well sample the number of SG chaotic attractors, $N_0$ is selected such that $\gamma_1 N_0$ ends in previously identified chaotic SG attractors where $\gamma_1\approx0.1$.
We indeed observed that the revisited SG attractors in this procedure occur almost exclusively once, which  strengthens our self-consistent  assumption that all SG attractors have similar basin of attractions. At the end of this step of the procedure $(1-\gamma_1)N_0$ distinct SG attractors are revealed. In the second step, we select $N_1$ new random initial conditions and calculate the fraction $\gamma_2$ of revisited $(1-\gamma_1)N_0$ recorded attractors similar to the previous step. The number of attractors is then estimated to be
\begin{equation}
\frac{\left(1-\gamma_1\right)N_0}{\gamma_2}.
\end{equation}
We expect this procedure to yield a lower bound for the number of SG attractors, since attractors with slightly larger basin of attractions are expected to be revealed with higher probability  in the first step of the algorithm. Figure 3 depicts the estimated number of SG chaotic attractors as a function of $N$ following the above procedure for $\epsilon=0.5$ and $0.9$. The estimation was carried out in the range of $N=[30,60]$, since for $N<30$ the distributions of $\Delta^{\alpha,\alpha}$ and $\Delta^{\alpha,\beta}$ (Fig. \ref{fig:hist}c-d), are significantly overlapped and for $N>60$ the number of attractors is too large to be reliably estimated in a reasonable computational time. Results indicate that the number of SG chaotic attractors scales exponentially with the size of the system, $e^{AN}$ with $A\sim 0.19$.
It seems that the prefactor $A$ slightly increase with $\epsilon$ as the local dynamics term in \eqref{dynamics} is weakened.
As a conclusion, a SG phase with exponentially many chaotic attractors is found.

\begin{figure}[t]
\includegraphics[width=0.5\textwidth]{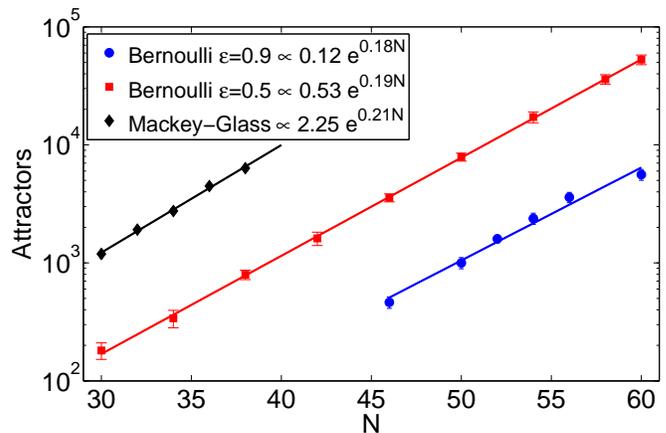}
\caption{(color online). The estimation of the number of SG chaotic attractors as a function of $N$ for $\epsilon=0.5$ (red) and $\epsilon=0.9$ (blue). For each $N$ and $\epsilon$ the average number of SG chaotic attractors and the standard deviation  were derived from at least $10$ pseudo-inverse systems.
Results for the number of chaotic attractors for the Mackey-Glass  and $\alpha=0.5$ (black)}
\label{fig:numOfAttractors}
\end{figure}

\subsection{Spin-Glass Phase}\label{sec: Spin-Glass Phase}
\begin{figure}[t!]
\includegraphics[width=0.5\textwidth]{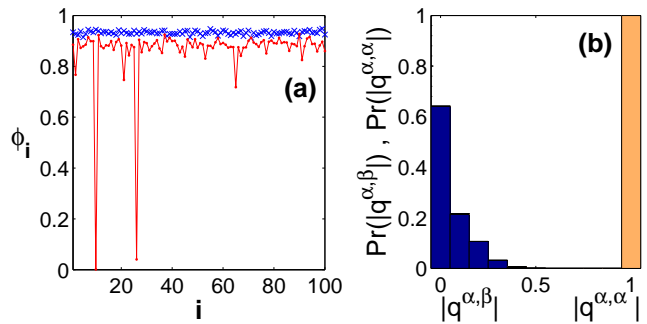}
\caption{
(color online).
(a) Freezing of each one of the $N=100$ units in a trajectory with an overlap close to unity with a patterns (blue-upper curve) and in a SG chaotic attractor (red-lower curve).
(b) The overlap $q^{\alpha,\beta}$, identifying a relative direction of freezing  between two chaotic SG attractors (dark blue), and between two vastly separated ($10^4\tau$ steps) time windows along a trajectory of a given SG chaotic trajectory $q^{\alpha,\alpha}$. The probability density $Pr(q^{\alpha,\beta})$ was derived  from $100$ chaotic SG attractors, whereas $Pr(q^{\alpha,\alpha})$ was constructed from  $100$ windows of length $100\tau$ (light orange).
}
\label{fig: freezing}
\end{figure}

A conventional method to identify a SG phase is a freezing of each degree of freedom , $X_i$, which cannot be directly deduced from the freezing   estimated by the Manhattan distance, \eqref{Manhattan distance}.
To measure the freezing of $X_i$ we introduce a quantity
\begin{equation}\label{freezing}
\phi_i=\frac{\langle X_i \rangle^2}{\langle X_i^2 \rangle}
\end{equation}
where $\langle ~ \rangle$ stands for an average over time steps.
Eq. \eqref{freezing} is analogous to the Edwards-Anderson order parameter,
$q_{EA}= \frac{1}{N} \sum_i \phi_i$ \cite{3,6}.
Figure \ref{fig: freezing}a indicates that along a trajectory inside a basin of attraction of a given pattern (blue curve) all units are almost completely frozen, $\phi_i \sim 0.95$.
Note that for a uniform distribution of $X_i$ with a given sign, e.g. $X_i \in [0,1]$, one can verify that $\phi_i=0.75$; hence $\phi_i \sim 0.95$ indicates  a freezing beyond the preservation of the sign of each $X_i$.
Indeed, a closer look at the distribution of each $X_i$ indicates that the distribution has considerably narrowed, e.g. mostly within the limited range $X_i \in [0.5,1]$.
A typical freezing within a SG chaotic attractor is depicted by the red curve of Fig. \ref{fig: freezing}a.
Almost all the units are highly frozen; however, a slight degradation in the freezing is observed in comparison to the blue curve.

The relative freezing direction between two chaotic SG trajectories is defined by
\begin{equation}\label{q}
q^{\alpha,\beta}= \frac{ \langle {\bf X}^\alpha \rangle \cdot \langle {\bf X}^\beta \rangle}
{\|\langle {\bf X}^\alpha \rangle\| \ \|\langle {\bf X}^\beta \rangle\|}
\end{equation}
similarly to the definition of the overlap between two thermodynamic states of a SG system \cite{3,6} and  $q^{\alpha,\beta}\in[-1,1]$ .
It is expected that the absolute overlap between two different SG chaotic trajectories, $|q^{\alpha,\beta}|,~\alpha \ne \beta$ is close to zero whereas the overlap between two vastly separated time windows along a trajectory of a given SG chaotic trajectory, $|q^{\alpha,\alpha}|$, is close to unity. Figure \ref{fig: freezing}b such a behavior where a substantial gap between the distributions of $|q^{\alpha,\beta}|$ and $|q^{\alpha,\alpha}|$ is observed, which is in agreement also  with Fig. \ref{fig:hist}c-d.

A crucial question is whether the SG chaotic attractors are closely related to the metastable states of the pseudo-inverse Ising spin system, Eqs. (\ref{Hamiltonian},\ref{pseudo-inverse}), with or without the diagonal terms.
This analogy is stimulating since the number of metastable states for the infinite-range SG system also scales   exponentially with $N$ and with $A\sim 0.2$ \cite{6,6a}.
We examined this question by clipping the sign of the frozen units in a SG attractor, $S_i^0 =sign\left(\langle X_i \rangle\right)$, and then evolved the pseudo-inverse system, eq. \eqref{Hamiltonian}, using zero temperature Monte-Carlo dynamics to the nearest metastable state with the Hamiltonian scenario, $W_{ii}=0$, or with $W_{ii} \ne 0$ and  with long dynamics and obtained a configuration $\{S_i^t\}$.
In both cases the overlap $\frac{1}{N} \sum_i S_i^0S_i^t$ was found to be close to zero, indicating no simple interplay between metastable  states of the energy surface, eq. \eqref{Hamiltonian}, and the chaotic SG attractors.
The lack of correlation was also obtained in the reverse scenario where the chaotic dynamics, $\{X_i\}$, was initiated by a metastable state of eq. \eqref{Hamiltonian}.

\section{Continuous Time Systems}\label{sec: Continuous Time Systems}
Finally we report on the extension of the exponentially many chaotic attractors to a network of coupled
differential equations, the Mackey-Glass (MG) equation \cite{14}, originated
from  physiological control systems.
This equation was also found to have many applications in hematology, cardiology, neurology and psychiatry \cite{MG2,MG3}.
The counterpart network version of the MG equation is given for unit $i$, for instance, by
\begin{equation}\label{eq: Mackey-Glass}
\frac{d}{dt}X_i(t)=-\gamma X_i(t) + \beta \sum_{j=1}^{N} W_{ij}\frac{X_j(t-\tau)}{1+\left|X_j(t-\tau)\right|^c}
\end{equation}
where in the notations of eq. \eqref{dynamics}
\begin{equation}
\mathcal{F}(x)=-\gamma x(t)
\end{equation}
and
\begin{equation}
\mathcal{G}(x)=\frac{x(t-\tau)}{1+\left|x(t-\tau)\right|^c}
\quad.
\end{equation}
As before, $N$  stands  for  the number of units in the  network  and the matrix $W_{ij}$ is given by eq. \eqref{pseudo-inverse}.
The parameters used in our simulations are  $\beta=2$, $\gamma=1$, $\tau=2$, $c=10$.

Using same arguments as in \eqref{state vector}, also in this case $\delta_\bot^t$ decreases exponentially to zero.
Hence, the stability of the Mattis states as well as the size of the basin of attractions can be estimated as in section \ref{sec: Mattis}. Similarly, the number of SG chaotic attractors can be estimated using the same procedure used in section \ref{sec: Number of Attractors}.
Results for the number of chaotic SG attractors for $\alpha=0.5$ are depicted in Fig. \ref{fig:numOfAttractors}, indicating again an exponential scaling with the network size, $N$.

\section{Conclusions}\label{sec: Conclusions}
We investigated discrete and continuous time networks of chaotic units with delayed interactions.
It has been shown that the networks function as an associative memory with a macroscopic basin of attraction (Fig. \ref{fig:basin}).

Moreover, a SG phase which is characterized by an exponential number of attractors (Fig. \ref{fig:numOfAttractors}) is identified, where
two nearby trajectories within a given attractor diverge from each other with a positive Lyapunov exponent \eqref{LE}.

It is worthwhile to note that the slope of the exponential number of chaotic attractors is very similar for both the Bernoulli map and the Mackey-Glass system, $\approx 0.19$ and $\approx 0.21$, respectively, which might indicate a universal behavior.
Both the Bernoulli map and the Mackey-Glass equation are sign preserving, even in the generalized forms eqs. \eqref{bernoulli map} and \eqref{eq: Mackey-Glass}.
An open question is whether multiple attractors and  a similar SG behavior can be found in a chaotic dynamics with the lack of sign preservation.

\end{document}